\documentclass[journal=jctcce, layout=twocolumn]{achemso}

\usepackage[T1]{fontenc} 

\usepackage{amsmath,color, tikz}
\usetikzlibrary{matrix}
\usepackage{amssymb}
\usepackage{amsfonts}
\usepackage{amsthm}
\usepackage{mathtools}
\usepackage{comment}
\usepackage{hyperref}
\hypersetup{
	colorlinks=true,
	citecolor=blue,
	linkcolor=red,
	urlcolor = blue
}
\usepackage{cleveref}
\usepackage{bm}

\usepackage{pythonhighlight}

\usepackage{algorithm}
\usepackage{algpseudocode}
\usepackage{dsfont}

\usepackage{bbold}
\usepackage{chemformula}
\usepackage{siunitx}

\DeclarePairedDelimiterX\braket[2]{\langle}{\rangle}{#1 \delimsize\vert #2}
\DeclarePairedDelimiterX\braket3[3]{\langle}{\rangle}{#1 \delimsize\vert #2 \delimsize\vert #3}

\usepackage{accents}
\usepackage{xspace}

\usepackage{array}
\usepackage{longtable}
\newlength{\topicwidth}
\usepackage{fancyvrb}
\usepackage{pdfpages}

\newcommand{\red}[1]{{\color{black} #1}}

\newcommand{\FermiLink}{\textsc{FermiLink}\xspace}

\usepackage{listings}
\usepackage{xcolor}
\usepackage{multirow}
\usepackage{xcolor}
\usepackage{pifont}

\newcommand{\highfidelity}{High-fidelity}

\newcommand{\expertpkgrow}[5]{%
  #1 & #2 & #3 & #4 & #5  \\
}
\newcolumntype{L}[1]{>{\RaggedRight\arraybackslash}p{#1}}

\usepackage{makecell}       

\definecolor{codegreen}{rgb}{0,0.6,0}
\definecolor{codegray}{rgb}{0.5,0.5,0.5}
\definecolor{codepurple}{rgb}{0.58,0,0.82}
\definecolor{backcolour}{rgb}{0.95,0.95,0.92}

\lstdefinestyle{mystyle}{
    backgroundcolor=\color{backcolour},   
    commentstyle=\color{codegreen},
    keywordstyle=\color{magenta},
    numberstyle=\tiny\color{codegray},
    stringstyle=\color{codepurple},
    basicstyle=\ttfamily\footnotesize,
    breakatwhitespace=false,         
    breaklines=true,                 
    captionpos=b,                    
    keepspaces=true,                 
    showspaces=false,                
    showstringspaces=false,
    showtabs=false,                  
    tabsize=2
}

\usepackage[T1]{fontenc}
\usepackage[utf8]{inputenc}
\usepackage{xcolor}
\usepackage[most]{tcolorbox}
\tcbuselibrary{breakable,skins,listings}

\usepackage{ragged2e}
\usepackage[most]{tcolorbox}
\definecolor{promptbg}{RGB}{245,245,245}

\newtcolorbox{llmprompt}{
  enhanced,
  breakable,
  colback=promptbg,
  colframe=black!55,
  coltext=black,
  boxrule=0.5pt,
  arc=1mm,
  left=1.5mm,right=1.5mm,top=1mm,bottom=1mm,
  halign=justify,
  before upper={\small\justifying\setlength{\parindent}{0pt}}
}

\makeatletter
\AtBeginDocument{\let\LS@rot\@undefined}
\makeatother

\author{Gang Meng}
    \affiliation{Department of Physics and Astronomy, University of Delaware, Newark, Delaware 19716, USA}

    \author{Andres Felipe Bocanegra Vargas}
    \affiliation{Department of Physics and Astronomy, University of Delaware, Newark, Delaware 19716, USA}

    \author{Xinwei Ji}
    \affiliation{Department of Physics and Astronomy, University of Delaware, Newark, Delaware 19716, USA}

    \author{Federico Garcia-Gaitan}%
	   \affiliation{Department of Physics and Astronomy, University of Delaware, Newark, Delaware 19716, USA}

    \author{Felipe Reyes-Osorio}%
	   \affiliation{Department of Physics and Astronomy, University of Delaware, Newark, Delaware 19716, USA}

    \author{Jalil Varela-Manjarres}%
	   \affiliation{Department of Physics and Astronomy, University of Delaware, Newark, Delaware 19716, USA}

    \author{Yafei Ren}%
	   \affiliation{Department of Physics and Astronomy, University of Delaware, Newark, Delaware 19716, USA}

    \author{Mohammadhasan  Dinpajooh}
    \affiliation
    {Physical and Computational Sciences Directorate, Pacific Northwest National Laboratory, Richland WA 99352, USA}

    \author{Branislav K. Nikoli\'{c}}%
	   \affiliation{Department of Physics and Astronomy, University of Delaware, Newark, Delaware 19716, USA}
        
	   \author{Tao E. Li}%
	   \email{taoeli@udel.edu}
	   \affiliation{Department of Physics and Astronomy, University of Delaware, Newark, Delaware 19716, USA}
	
	\title{\red{\FermiLink: An Agent Harness for Multidomain Scientific Simulations}}

\begin{document}

    \begin{tocentry}
    \centering
        \includegraphics[width=0.5\linewidth]{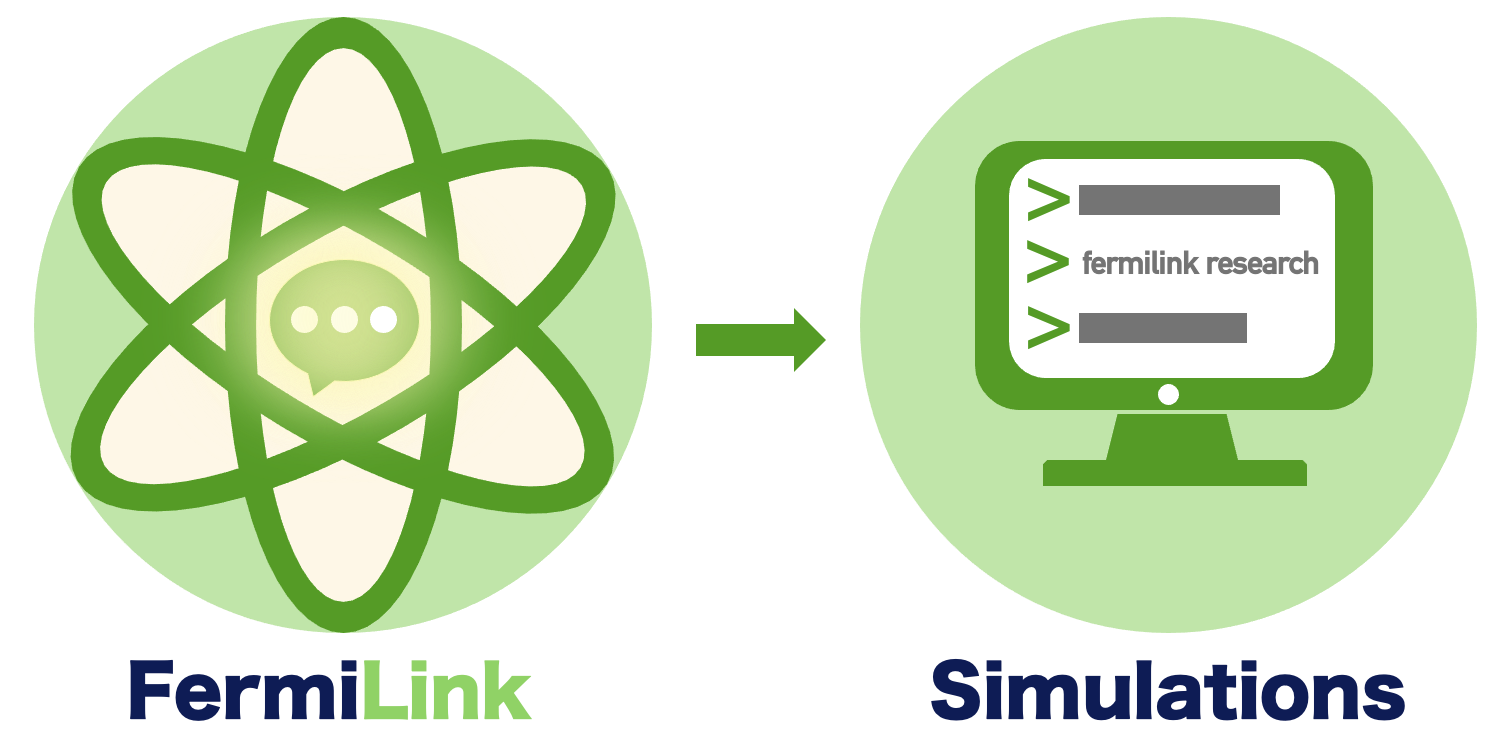}
    \end{tocentry}
    
    \begin{abstract}
    Artificial-intelligence (AI) agent \red{harnesses} have been developed for autonomous scientific simulations, but most current agent \red{harnesses are tested on} single or a small set of software packages.  Herein, \FermiLink, \red{an extensible open-source agent harness} is introduced for multidomain scientific simulations. Its key design principle is the separation of package knowledge bases from simulation workflows, so that simulation workflows in \FermiLink, from figure-level simulations to full-paper-level research on high-performance computing clusters, operate uniformly among supported packages via a four-layer progressive disclosure mechanism. \red{This agent harness is specifically optimized for high performance computing (HPC) clusters, enabling multi-day multi-step scientific simulations in the absence of human intervention.} Using OpenAI Codex as the agent provider, the capabilities of \FermiLink are demonstrated across approximately 50 scientific software packages spanning nine research domains from \red{chemistry} to  engineering. Systematic benchmarks on 132 real-world figure-level reproduction tasks with 44 packages show that \FermiLink reproduces 74 (56.1\%) of published figures with simulations, among which 30 \red{(22.7\%)}  achieve high-fidelity agreement and 35 reach qualitative agreement with the target figures. A smaller set of human expert-guided reproduction benchmarks  with 10 packages \red{in physics and chemistry} further highlights the importance of expert insights for improving the simulation fidelity.  Beyond reproduction, a single-blinded study demonstrates that \FermiLink can produce research-grade results on unpublished polariton physics problems when provided with sufficiently detailed research objectives. Overall, \FermiLink provides a scalable research infrastructure that  may accelerate the path from scientific questions to computational results across diverse domains.

    \end{abstract}

	\maketitle

    \setcitestyle{super}

    \section{Introduction}

    Computational simulations play a central role in modern scientific discovery~\cite{Dongarra2024,Barbatti2025,Sadybekov2023,Waintal2024,Post2005}. Very often, these calculations utilize different homegrown or large-scale open-source and commercial scientific software packages. Some of these packages provide well-structured tutorials and  documentation; however, many offer only limited usage examples beyond the released source code. As a result, mastering each computational package and efficiently  executing scientific simulations on high-performance computing (HPC) clusters remain major bottlenecks in modern research workflows.

    Large language model (LLM)-based artificial intelligence (AI) \cite{Brown2020GPT3FewShot,OpenAI2023GPT4,Chen2021Codex,Yang2024SWEAgent,Boiko2023LLMChem,M.Bran2024AugmentingLLM,Lu2024AIScientist,Gottweis2025AIcoscientist,Schmidgall2025AgentLab,Ramos2025LLMChem} technologies are beginning to revolutionize computational simulations in natural sciences. For instance, in theoretical chemistry, an AI chatbot was developed for performing first-principles solvation calculations \cite{Gadde2025ChatbotAssistedChemSci}. Very recently,  AI agent \red{harnesses} for classical molecular dynamics \cite{Campbell2026MDCrowMLST}, quantum chemistry \cite{Zou2025ElAgenteMatter},
    quantum dynamics simulations \cite{Gustin2025ElAgenteCuanticoArXiv} and high-energy physics~\cite{Schwartz2026} have been reported. In other computational fields,  agent \red{harnesses} have also been developed for automating workflows involving a single or a small set of computational packages~\cite{hu2026tritondft,Wang2025aDREAMS,Yao2025MultiAgent}. \red{Recent studies have further extended this direction toward autonomous generation of computational tools and self-evolution of agent skills\cite{huangCASCADECumulativeAgentic2026,zhangAgenteForjadorTaskDriven2026}.}

    However, this \red{domain-specific} approach has significant limitations. Connecting $N$ agent \red{harnesses} to $M$ scientific software packages demands up to $N \times M$ individual integrations. This combinatorial bottleneck may drastically limit the broader adoption of AI agents in computational research. More importantly, the rapid performance improvement of commercial LLM providers (such as OpenAI, Claude, and Google Gemini) requires swift adjustment of agent \red{harnesses} for adapting to the LLM performance change. As such, it will spread  tremendous human efforts  for maintaining and developing package-specific agent \red{harnesses}.

\begin{figure*}
  \centering
  \includegraphics[width=0.9\textwidth]{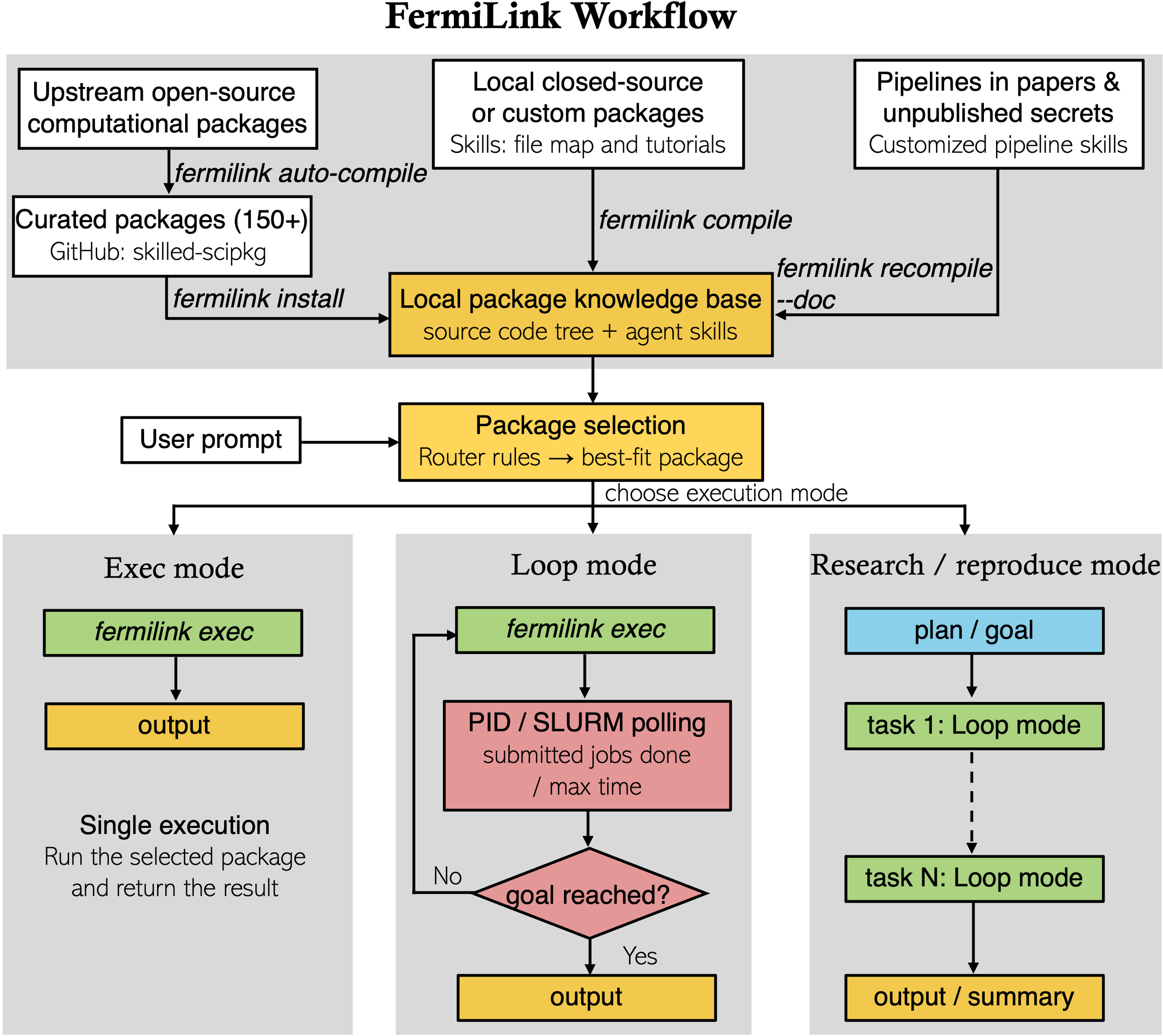}
  \caption{Design of the \FermiLink agent \red{harness}. \FermiLink dynamically loads the most suitable package knowledge base to respond to the user's request. Three major workflows are supported in \FermiLink: \texttt{exec}, \texttt{loop}, and \texttt{research/reproduce} for processing computational simulations at different scopes. As the package knowledge bases are segregated from simulation workflows, \FermiLink provides a unified agent \red{harness} for multidomain scientific simulations.}
  \label{fig:workflow}
\end{figure*}

Here, we introduce \FermiLink, an extensible  agent \red{harness} for multidomain scientific simulations\red{, complementary to existing agent harnesses for scientific simulations in each domain.} As shown in Figure~\ref{fig:workflow}, by separating simulation workflows from package knowledge bases, \FermiLink is uniformly applicable to computational packages across multiple domains. Workflows at different levels have been designed for different purposes, ranging from small-scale laptop simulations to long-duration (days or longer) simulations on HPC clusters and multi-task research-level simulations. On the software knowledge base, \FermiLink provides a forward-thinking design principle: it exposes the entire package source code tree plus a pre-compiled agent skills layer for agent reasoning\red{, as modern LLM models become increasingly capable for reasoning on large-scale code bases.}
By incorporating more than 150 built-in software knowledge bases \cite{skilled-scipkg-repositories} and transferring source-grounded domain knowledge of simulations to the agent via a four-layer progressive disclosure mechanism (Sec.~\ref{sec:methods}), \FermiLink \red{\ functions as an agent harness for efficiently reasoning and executing long-duration simulation workflows from research objectives defined by humans.
We believe that this agent harness may} offer a  scalable research infrastructure for multidomain scientific simulations. 

\section{Methods}\label{sec:methods}

The key design principle of \FermiLink is the segregation of  package knowledge bases and simulation workflows. This separation is inspired by the commonalities and differences inherent in scientific computing. For example, almost all scientific simulations involve simulation pipelines utilizing structured input files on local machines or HPC clusters; by contrast, the detailed parameter settings and required computing resources may vary significantly across different domains. To uniformly support multidomain computational simulations, \FermiLink contains built-in  knowledge bases for more than 150 scientific packages and adopts a \textit{four-layer progressive disclosure mechanism} to selectively feed necessary information to commercially available LLMs. 

This four-layer progressive disclosure mechanism, as shown in Figure \ref{fig:workflow}, is constructed as follows. (\textit{i}) Upon the user's request, \FermiLink dynamically loads the most suitable package knowledge base for agent reasoning. (\textit{ii}) When the agent starts to reason and simulate, it is instructed to load an agent skills \cite{Ling2026AgentSkills} layer first. The lightweight agent skills layer contains highly compressed tutorials for using the package, as well as an informative file map of the source code tree. (\textit{iii}) According to this informative file map, the agent can efficiently load the most relevant files in the source code tree for processing the user's request. (\textit{iv}) Simulation pipelines from research papers or unpublished results can also be appended to the agent skills layer with a single command line setting in \FermiLink, so that this package can perform not only demonstrative simulations but also production calculations at the publication level. Hence, we name this agent framework \textsc{F}idelity-\red{\textsc{E}nhanced} \textsc{R}etrieval for \textsc{M}odular \textsc{I}ntegration (\textsc{FERMI})-\textsc{Link}, which connects natural-language requests to faithful, source-grounded simulation pipelines through progressive disclosure. 

    To accommodate simulations at different scopes, as demonstrated in Figure~\ref{fig:workflow}, \FermiLink delivers with three major computational workflows. While the \texttt{exec} mode is designed for short-duration simulations, the \texttt{loop} mode connects iterative agent reasoning with simulation monitoring for PID and SLURM jobs, thus providing robust support for long-duration simulations on both workstations and HPC clusters. The \texttt{research/reproduce} mode is further intended for multi-task simulations at the scope of a full research paper. 

    The \FermiLink agent framework utilizes state-of-the-art coding agents (supporting OpenAI Codex, Claude Code, Gemini CLI\red{, and a wide range of open-weight LLMs via OpenCode}) for reasoning and local operations, while \FermiLink itself focuses on the construction of package knowledge bases and development of multiple simulation-specific workflows \red{for robust multi-day multi-step scientific simulations. For example, because individual scientific simulation tasks may require days or weeks for computation, the \FermiLink harness controls the monitoring of HPC jobs and handles the completed or failed simulations to coding agents for further reasoning. By contrast, bare coding agents may simply stop reasoning after submitting the HPC jobs or polling the SLURM status and reasoning too frequently.}

    \FermiLink provides a set of command-line tools for experienced users, as well as access to other AI agents. Additionally, \FermiLink  supports a Web-based user interface for a ChatGPT-like experience plus remote controlling using popular messaging apps (SI Sec. \red{VI}). For example, users can utilize Telegram on their cellphones to communicate with many copies of \FermiLink agents hosted on HPC clusters for performing various large-scale HPC calculations in parallel. The unified short-term/long-term memory mechanism further allows \FermiLink remembering  setups and pitfalls in previous calculations, a feature that is particularly appealing for long-term research projects. Beyond these features, to facilitate users in evaluating the validity and fidelity of simulations~\cite{Post2005,Hatton1997,Williams2020},  \FermiLink is designed to always provide uncertainty information and confidence gaps of the simulations.

    \red{All reported simulations in the main paper were conducted between February and April 2026 using \texttt{gpt-5.3-codex} under reasoning effort \texttt{xhigh}. Because \texttt{gpt-5.3-codex} is a proprietary model whose behavior may change across model updates, the reported results may not be fully reproducible beyond this evaluation period.
    Calculations were performed through two ChatGPT Pro monthly subscriptions.
    }
    Detailed usage of the \FermiLink agent framework is provided at GitHub \url{https://github.com/TaoELi/FermiLink}. 

\section{Results}

We demonstrate the key capabilities of \FermiLink through three sets of examples.
These examples not only showcase the use of \FermiLink for reproducing published results in multidomain scientific simulations, but also highlight a practical workflow for \red{autonomously executing complex research-scale scientific simulations.}

\subsection{Reproducing figure-level results in multiple scientific domains}

To examine whether the current design of \FermiLink is capable of multidomain scientific simulations, we assembled a benchmark spanning \textit{44} scientific packages  drawn evenly from the currently available package knowledge bases (150+) in \FermiLink. For each package, we choose three computational tasks, each for reproducing one figure in a published paper using this package. In total, 132 different figure-level tasks are conducted using the \FermiLink \texttt{loop} mode (Figure \ref{fig:workflow}). \red{Rather than focusing on the performance difference of various LLM models, this study  serves as a system-level benchmark using diverse scientific tasks with a uniform LLM model (\texttt{gpt-5.3-codex}).}

For these tasks, a uniform prompt is given as follows:
\begin{llmprompt}
\texttt{
Use <pkg-id> package to reproduce <figure> in the paper <paper-url> using identical parameters as the paper. Install this package locally if you cannot find it installed. Record the path to the locally installed package to memory.md so future jobs do not need to reinstall the same package.
}
\end{llmprompt}
\red{For minimizing potential selection bias, the benchmark packages and corresponding references were selected automatically by an independent AI agent instead of being manually chosen by the authors. Importantly, the selection was performed without prior assessment of task difficulty, and no task was discarded on the basis of   subsequent performance of \FermiLink.}

\begin{figure*}
  \centering
  \includegraphics[width=1.0\textwidth]{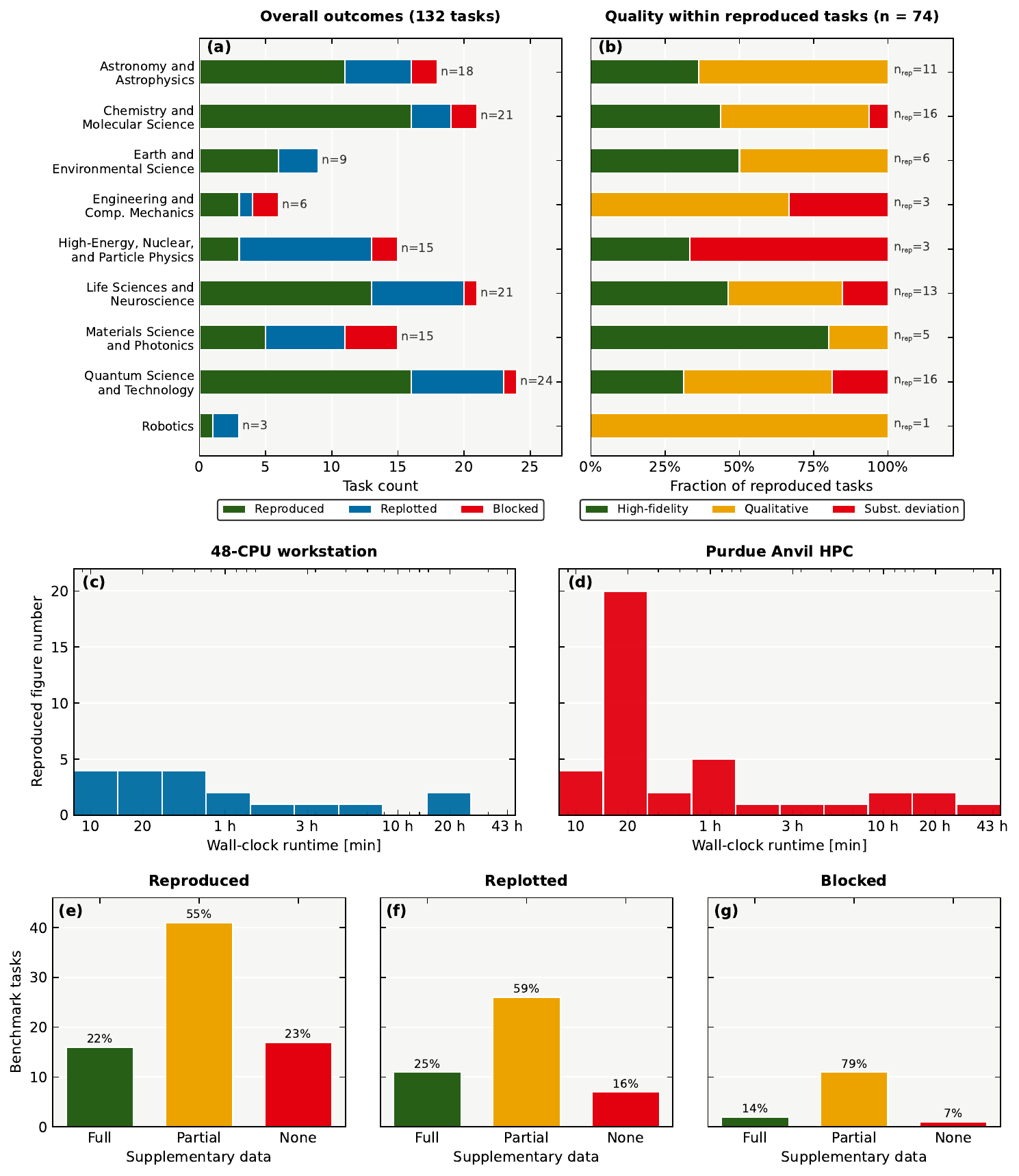}
  \caption{Summary of the 132 figure-level reproduction tasks in Table~\red{S1 of the  Supporting Information (SI)}. (a) Outcome distribution across nine scientific domains: Reproduced (simulation rerun with the target package, green), Replotted (figure generated from published or extracted data without new simulation, blue), and Blocked (figure not produced, red). (b) Reproduction quality among the 74 Reproduced tasks. (c, d) Wall-clock runtime distributions for the Reproduced tasks on (c) a 48-CPU workstation and (d) the Purdue Anvil HPC cluster. Runtimes include the full agent workflow from package installation through post-processing. (e--g) Supplementary-data availability distributions for the Reproduced, Replotted, and Blocked outcome categories, respectively. }
  \label{fig:benchmark_summary}
\end{figure*}

As analyzed in Figure \ref{fig:benchmark_summary}a, the 132 figure-level tasks (\red{Table~S1 of the SI}) are classified into three outcomes: \textit{Reproduced} (56.1\%), where \FermiLink reruns the simulation using the target package and generates the figure from new computational results; \textit{Replotted} (33.3\%), where no new simulation is performed and the figure is generated from released data, or simply values extracted from published figures; and \textit{Blocked} (10.6\%), where the final figure cannot be produced. Among the 74 reproduced tasks with actual simulations (Figure~\ref{fig:benchmark_summary}b), 30 (40.5\%) achieve high-fidelity agreement with published results, 35 (47.3\%) show qualitative agreement, and 9 (12.2\%) exhibit substantial deviation. The overall high-fidelity reproduction rate across all 132 tasks is 22.7\%. \red{Note that the reproduction quality was independently assessed by two human evaluators, with a third evaluator consulted to resolve disagreements. The detailed evaluation protocol is explained in Sec. I of the SI.}

Chemistry and quantum sciences contributed the largest shares of reproduced tasks (Figure~\ref{fig:benchmark_summary}a). Runtime distributions (Figures~\ref{fig:benchmark_summary}c and d) show that simulations span from minutes to over 24 hours, demonstrating the framework's ability to sustain long-running computations at HPC or workstations. \red{Interestingly, the HPC runtime performance is dominated by approximately 20 minutes. This peak corresponds to a single-session Codex reasoning, in which the agent may successfully reproduce the results with minimal computation or simply fail to accomplish the task due to a lack of information.} As shown in the supplementary data availability analysis in Figures~\ref{fig:benchmark_summary}e--g, the blocked tasks are overwhelmingly associated with incomplete supplementary data, confirming that data availability remains a critical determinant of reproducibility.

The prevalence of replotted tasks (33.3\%) reveals an important behavioral pattern: When simulation inputs are unavailable, the agent defaults to reproducing the visual output rather than reporting failure. While it may be acceptable to replot the figures using published supplementary data, we also witness the agent behavior on extracting pixel data directly from published figures, which is functionally copying. This shortcut-seeking behavior underscores the need for process-level validation \red{when deploying AI agents for scientific simulations; after all, simply output-level comparison may not be sufficient.}

As shown in Figures \ref{fig:benchmark_summary}c and d, because most the simulation jobs are finished within a few hours, one may wonder whether bare coding agents without the \FermiLink \texttt{loop} harness may also perform similarly as \FermiLink. To reveal the role of the \FermiLink harness, we then repeat the 21 Chemistry and Molecular Science tasks using bare Codex with the same \texttt{gpt-5.3-xhigh} model and task prompts used for \FermiLink (Table S2 of the SI). Overall, \FermiLink achieves significant improvements in the final outcome for five tasks compared with bare Codex. Through analysis (see also Table S3 of the SI), these improvements are mainly observed for workflows involving additional packages, package-specific implementation details, or online available source data that could lead the agent to replot instead of rerunning simulations. Clearly, the design of the \FermiLink harness, including the package/environment preparation and the coordination of multi-package scientific simulations, appears important for improving the agent performance.

\subsection{Reproducing scientific publications with expert insights}

\begin{table*}[t]
\centering
\begingroup
\scriptsize
\setlength{\tabcolsep}{4pt}
\renewcommand{\arraystretch}{1.05}
\caption{\textbf{Selected expert-guided \FermiLink tests beyond the zero-expert benchmark.}}
\label{tab:expert_guided_tests}
\begin{tabular}{@{}L{0.13\textwidth}L{0.15\textwidth}L{0.08\textwidth}L{0.12\textwidth}L{0.45\textwidth}@{}}
\hline
\textbf{Package} & \textbf{Task} & \textbf{Mode} & \textbf{Reprod. Quality} & \textbf{Notes} \\
\hline

\expertpkgrow{\texttt{Psi4}~\cite{pkg_psi4}}{Figures.~3 \& 4~\cite{paper_expert_psi4_1}}{\texttt{loop}}{High-fidelity}{A wider range of orbital energies was explored than those reported in Ref.~\citenum{paper_expert_psi4_1}. Report at SI Sec. I\red{V}.A.}

\expertpkgrow{\texttt{CP2K}~\cite{pkg_cp2k}}{Figure~5~\cite{paper_expert_cp2k_1}}{\texttt{loop}}{High-fidelity}{Only the results for the number of beads less than 256 were reproduced to reduce the computational cost. Report at SI Sec. I\red{V}.A.}

\expertpkgrow{\texttt{RMG-Py}~\cite{pkg_rmg_py}}{Figure~4~\cite{paper_expert_rmg_py_1}}{\texttt{loop}}{High-fidelity}{Report at SI Sec. I\red{V}.A.}

\expertpkgrow{\texttt{TeNPy}~\cite{pkg_tenpy}}{\makecell[l]{Figure~3~\cite{paper_expert_tenpy_1}}}{\texttt{exec}}{Qualitative}{Finite-size deviations observed. This was caused by a smaller system size specified in the prompt. Report at SI Sec. I\red{V}.B; below the same.}
\expertpkgrow{}{\makecell[l]{Figure~4~\cite{paper_expert_tenpy_2}}}{\texttt{exec}}{High-fidelity}{Minor finite-size effects.}

\expertpkgrow{\texttt{NESSi}~\cite{pkg_nessi}}{Figure~5~\cite{paper_expert_nessi_1}}{\texttt{exec}}{High-fidelity}{High frequency features are smeared due to attempts to avoid longer, more expensive calculations. Report at SI Sec. I\red{V}.C.}

\expertpkgrow{\texttt{QuTiP}~\cite{pkg_qutip}}{Figure~7~\cite{paper_expert_qutip_1}}{\texttt{exec}}{Qualitative}{There are difficulties converging in the low-temperature regime, where HEOM has intrinsic limitations. Report at SI Sec. I\red{V}.C.}

\expertpkgrow{\texttt{Kwant}~\cite{pkg_kwant}}{Figure~4~\cite{paper_expert_kwant_1}}{\texttt{exec}}{Qualitative}{Avoided crossings are sharper, as well as showing a slight overall shift in some features, possibly due to finite-size effects. Report at SI Sec. I\red{V}.D.}

\expertpkgrow{\texttt{MEEP}~\cite{pkg_meep}}{\makecell[l]{Figure 1c~\cite{christopher2012singular}}}{\texttt{loop}}{Qualitative}{ Electric field distribution mismatch around the edges and centers of the cubes. Report at SI Sec. I\red{V}.E; below the same.}
\expertpkgrow{}{Figures 2 \& 6 (btm.)~\cite{zaplotnik2023photonic}}{\texttt{loop}}{High-fidelity}{}
\expertpkgrow{}{\makecell[l]{Figure 2 ~\cite{ochiai2001dispersion}}}{\texttt{exec}}{High-fidelity}{Axis values are divided by 2.}
\expertpkgrow{}{Figure 1 (btm.)~\cite{hsu2013bloch}}{\texttt{loop}}{High-fidelity}{Photonic structure extended along the \textit{y}-axis.}

\expertpkgrow{\texttt{LAMMPS}~\cite{pkg_lammps}}{\makecell[l]{Figure 2\cite{paper_pmf_2025}}}{\texttt{reproduce}}{High-fidelity}{Run for 2 hours on 64 MPI ranks. Report at SI Sec. I\red{V}.F.}

\expertpkgrow{ \texttt{QuTiP}~\cite{pkg_qutip}}{\makecell[l]{Figures 2 \& 3 \cite{Grochowski2025}}}{\texttt{reproduce}}{\highfidelity}{Telegram remote control of HPC. Run for 12 hours. Report at SI Sec. I\red{V}.G,H.}

\expertpkgrow{modified \texttt{i-PI}/\texttt{LAMMPS}~\cite{liCavityMolecularDynamics2020}}{\makecell[l]{All figures \cite{liCavityMolecularDynamics2020}}}{\texttt{reproduce}}{\highfidelity}{Telegram remote control of HPC. Run for ten days. Report at SI Sec. I\red{V}.I,J.}
\hline
\end{tabular}
\endgroup
\end{table*}

While the above reproduction benchmarks rely on a one-shot prompt with zero human expert insights, some of the authors have also employed \FermiLink to perform a smaller set of reproduction tests (containing 10 scientific packages) in their specialized research fields using iterative conversations with the agent. As summarized in Table~\ref{tab:expert_guided_tests}, expertise in the field can greatly improve the fidelity for reproducing the simulation results, as the user can identify potential gaps more easily.

For instance, in the \texttt{QuTiP} package\cite{pkg_qutip} for open quantum system dynamics (SI Sec. I\red{V}.C.),  properly reproducing previously published results via hierarchical-equations-of-motion (HEOM) algorithm \cite{paper_expert_qutip_1} with \texttt{QuTiP} can only be achieved by identifying a factor of two difference in the definition of the environmental spectral density function in the manuscript versus \texttt{QuTiP} documentation.  After all, \FermiLink is designed to follow the guidelines of the source code tree (or package knowledge base) faithfully, so any internal conflicts between the manuscript and the source code tree may lead to incorrect reproduction of the paper.

Apart from the intrinsic conflicts between the documentation and publications, the large computational cost may also prohibit the efficient reproduction of the figures, such as many of the blocked calculations in \red
{Table~S1 of the SI}. However, with human expertise, by deliberately avoiding running expensive calculations and instead using reduced but still scientifically meaningful parameters, high-fidelity reproduction can still be partially achieved. For instance,  with  \texttt{CP2K} simulations\cite{pkg_cp2k} of \textit{ab initio} path-integral molecular dynamics (SI Sec. I\red{V}.A.),\cite{paper_expert_cp2k_1} we can avoid benchmarking a large number of path-integral beads and sample only a smaller number of trajectories than the manuscript, yet still recover quantitatively similar results.

Two final examples in Table~\ref{tab:expert_guided_tests} use the \FermiLink \texttt{reproduce} mode to successfully reproduce all the key data figures in full research papers. In both cases, due to the short-term/long-term memory mechanism of \FermiLink, once the initial figures are successfully reproduced, the agent can reuse intermediate outputs and bypass the previous pitfalls, thus moving forward at a faster pace. These paper-scale studies also highlight the current bottleneck of \FermiLink-enabled  computational simulations. \red{As one session of agent reasoning can usually be done within half an hour, the main delays may come from the computational cost of scientific simulations  and the restriction of HPC resources (such as multi-day simulations on many HPC nodes)}. 

\red{The results in expert-guided \FermiLink simulations indicate that, beyond our conclusion for Figure \ref{fig:benchmark_summary}}, the capacity of \FermiLink for sustaining long-duration (days or longer) multi-task simulations on HPC environments showcases its advantages over bare coding agents.

\begin{figure*}
  \centering
  \includegraphics[width=1\textwidth]{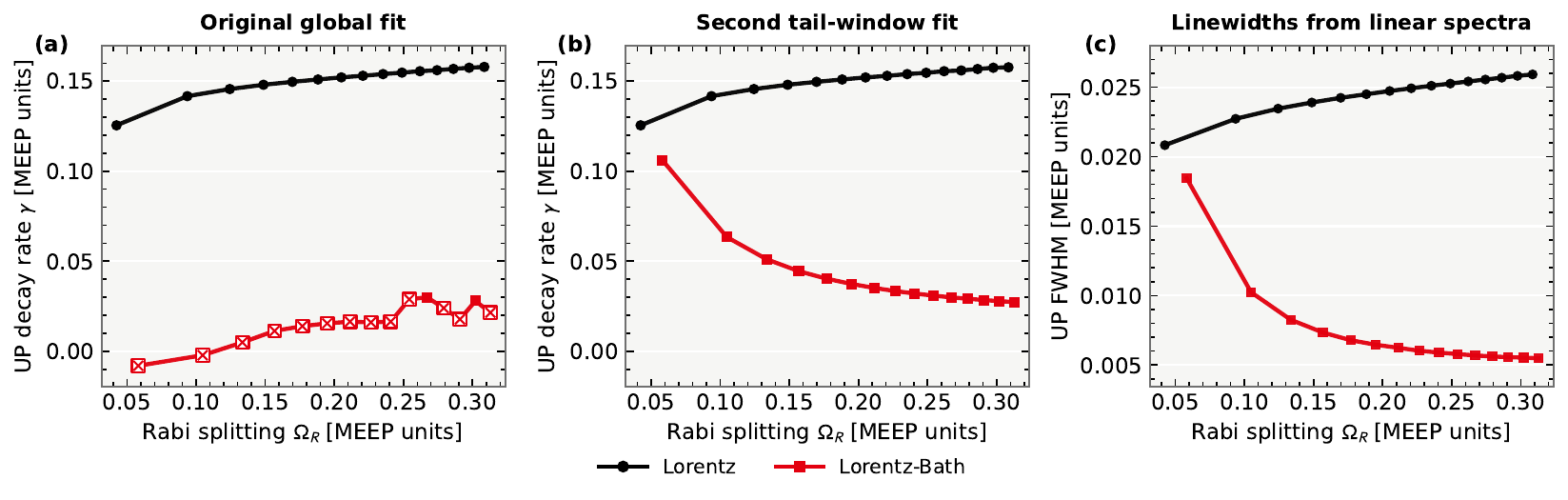}
  \caption{Comparison of calculated UP decay rates versus Rabi splitting for single-blinded simulations with the \texttt{FDTDBATH-MEEP} package. (a) First-attempt decay rates obtained by fitting the entire EM energy relaxation dynamics within the cavity. (b) Second-attempt  decay rates obtained by fitting only the tail-window of the EM energy relaxation dynamics. (c) UP linewidths extracted from linear spectroscopy. The agent rejects the results in part (a), as the Lorentz-Bath values (red) disagree with the linear-response linewidths in part (c), and instead focuses on fitting the tail-window of the EM energy relaxation dynamics [part (b)], which excludes the impact of EM energy accumulation from the narrow-band pulse excitation.}
  \label{fig:interesting_pattern}
\end{figure*}

\subsection{Combined \texttt{reproduce}/\texttt{research} workflows for \red{AI-assisted} scientific research: A single-blinded test}

Beyond reproducing known results, we then ask a more challenging question: {\em Can the \FermiLink framework execute a pre-specified computational research plan}? To explore this possibility, we design a single-blinded experiment around the \texttt{FDTDBATH-MEEP} package \cite{Li2025FDTDBath}, a modified version of the widely used \texttt{MEEP} package \cite{pkg_meep} for finite-difference time-domain (FDTD) simulations of classical electromagnetism. In addition to the capabilities of the standard \texttt{MEEP} package, this revised code implements a novel FDTD-Bath algorithm \cite{Li2025FDTDBath} for simulating condensed-phase polaritonics \cite{Ribeiro2018,Mandal2023ChemRev,Ruggenthaler2023}. Compared to the standard FDTD approach, the FDTD-Bath algorithm replaces the dissipation terms of the dielectric functions by the coupling to explicit bath oscillators, thus providing a more realistic description of EM fields interacting with molecules and materials.   Using this extended framework, a postdoctoral researcher has previously spent approximately two months generating unpublished results on the roles of bath anharmonicity and noise in polariton formation, and on the visualization of molecular dark-state dynamics under strong coupling in realistic two-dimensional optical cavities. These studies rely on several newly implemented features in \texttt{FDTDBATH-MEEP} for which no relevant online documentation or tutorial is available.

The single-blinded test proceeds as follows (SI Sec. \red{V}). The agent skills layer for \texttt{FDTDBATH-MEEP} includes the skill required to reproduce published FDTD-Bath results \cite{Li2025FDTDBath}. After reproducing Ref.~\citenum{Li2025FDTDBath} via the \texttt{reproduce} mode, we establish the correct computational environment for simulations. Then, we provide the \texttt{research} mode of \FermiLink with only a \texttt{goal.md} file containing the scientific objectives and the expected figure list (using the command \texttt{fermilink research goal.md}). Apart from the skills needed to efficiently locate the relevant source code, the agent is not given documentation or human-written instructions for using the advanced FDTD-Bath features required in this study, such as how to include bath anharmonicity and stochastic noise or visualize the dark-state dynamics, nor is this information available online.

Within 24 hours of iterative reasoning and simulation under the \texttt{research} mode, \FermiLink generates a research report that reproduces all the major scientific findings of the unpublished study, including seven multi-panel figures. Among all simulation results, one particular interesting self-reflection behavior pattern of the agent is worth noting. 

Following the guideline in \texttt{goal.md}, \FermiLink performs a series of simulations and plots the upper polariton (UP) decay rates versus Rabi splitting (Figure \ref{fig:interesting_pattern}a) by fitting the electromagnetic (EM) energy relaxation dynamics inside the cavity after narrow-band Gaussian-pulse excitation of the UP. After recognizing that most of the FDTD-Bath results (red crosses) disagree with the trend of the linear-response UP linewidths (Figure \ref{fig:interesting_pattern}c), the agent refuses to report Figure \ref{fig:interesting_pattern}a; instead, it tries again by fitting only the tail window of the EM energy relaxation dynamics, in which the effect of EM energy accumulation due to the incident pulse excitation becomes negligible. \red{In contrast, bare Codex followed the calculations specified in \texttt{goal.md} and accepted the initially obtained UP-state decay rates, resulting in an incorrect dependence on Rabi splitting for the Lorentz-Bath model (Sec. V.D of the SI).} Since providing consistent UP decay rates and linewidths is not required in \texttt{goal.md}, \red{this comparison highlights FermiLink’s capacity to detect physically inconsistent results and adapt its analysis strategy through self-correction}.

We emphasize that successfully generating multi-task simulation results is aided by the fact that we already knew which parameter regimes are scientifically relevant and which figures should be produced. Of course, without  prior knowledge of this important information, iterations of report generation and research objective modification may be needed.

Nevertheless, this single-blinded study suggests that, once a sufficiently detailed scientific direction is specified, \FermiLink may produce research-grade results based on the given source code of the computational package using the combined \texttt{reproduce}/\texttt{research} workflows, even in the absence of external documentation or tutorials. This single-blinded study also showcases the necessity for exposing the package knowledge base (including the whole source code tree) for agent reasoning\red{, which is a} key feature of \FermiLink.

\section{Conclusion}

In summary, we have implemented  \FermiLink as \red{an extensible AI agent harness} for  scientific computational simulations and demonstrated its capabilities by applying it to numerous software packages (Tables~\red{S1 of the SI} and Table \ref{tab:expert_guided_tests}) across wide range of scientific disciplines. Due to the design principle of separating package knowledge bases and simulation workflows, \FermiLink enables \red{multi-day, multi-step} scientific simulations within the same agent framework. More importantly, this study suggests that \FermiLink can move beyond demonstration and function as a practical tool for massive reproduction of published simulation results, as well as for \red{executing research-scale computational workflows involving previously unexplored simulation tasks}.

\red{Despite these progresses, the limitations of this work are equally instructive. First, this work does not provide a comprehensive benchmark using a set of different LLM models or comparing to other published agent harnesses for scientific simulations. Due to a lack of this comparison, a quantitative measurement of the \FermiLink agent harness capacity cannot be established. Second, as the current coding agents such as Codex and Claude Code are becoming increasingly more capable of executing complicated workflows, swift and continuous adjustment of the \FermiLink harness for adapting to this change is needed.}

Broadly speaking, the benchmark of \FermiLink suggests that the near-term value of AI in scientific simulations, if properly designed, is the potential for taking over a substantial share of slow, repetitive work between a scientific question and practical simulation outcomes. Still, the human scientific expertise in each domain is needed, perhaps more urgently, for proposing detailed and practical simulation objectives and evaluating the validity of the simulation outcomes and their scientific importance\red{. After all,} the agent may seek shortcuts to achieve the final objective.

\begin{acknowledgement}
This material is based upon work supported by the U.S. National Science Foundation (NSF) under Grant No. CHE-2620630 (for the development of \FermiLink agent framework) and Grant No. CHE-2502758 (for polariton-related simulations). F.G.-G., F.R.-O., J.V.-M. and B.K.N. were additionally supported by NSF under Grant No. DMR-2500816. M.D. was additionally supported under FWP 85666, a U.S. Department of Energy (DOE), SC, Early Career Research Program award in the Basic Energy Sciences (BES), Chemical Sciences, Geosciences, and Biosciences (CSGB) Division, Condensed Phase and Interfacial Molecular Science (CPIMS) program (for applying \FermiLink to aqueous solutions).  This work used the Anvil HPC at Purdue University through allocation CHE260067 from the Advanced Cyberinfrastructure Coordination Ecosystem: Services \& Support (ACCESS) program \cite{Boerner2023}, which is supported by U.S. National Science Foundation grants \#2138259, \#2138286, \#2138307, \#2137603, and \#2138296.
\end{acknowledgement}

\section*{Data and Software Availability}
    The \FermiLink package used in this manuscript is available at Github \url{https://github.com/TaoELi/FermiLink}. Simulation data of this manuscript are archived in \url{https://www.taoeli.org/publications}.

\begin{suppinfo}
    Supporting information PDF file includes benchmark table corresponding to Figure \ref{fig:benchmark_summary};  21 Chemistry related reproduction jobs using bare Codex; Supporting reports for Table \ref{tab:expert_guided_tests}; Supporting data for the FDTD-Bath simulations; and Web UI and Telegram interface of \FermiLink.
\end{suppinfo}

    \providecommand{\latin}[1]{#1}
\makeatletter
\providecommand{\doi}
  {\begingroup\let\do\@makeother\dospecials
  \catcode`\{=1 \catcode`\}=2 \doi@aux}
\providecommand{\doi@aux}[1]{\endgroup\texttt{#1}}
\makeatother
\providecommand*\mcitethebibliography{\thebibliography}
\csname @ifundefined\endcsname{endmcitethebibliography}  {\let\endmcitethebibliography\endthebibliography}{}

    \end{document}